\documentclass[prl,aps,twocolumn,superscriptaddress]{revtex4}
\usepackage{psfrag,slashed,cancel,array,graphicx}
\usepackage[utf8]{inputenc}
\usepackage{mathtools}
\usepackage{mathrsfs}
\usepackage{multirow}
\usepackage{braket}
\usepackage{slashed}
\usepackage{booktabs}
\usepackage{slashed}
\usepackage{hyperref}
\usepackage{cleveref}
\usepackage{amssymb}

\hypersetup{pdftitle={},pdfcreator={},linkcolor=[rgb]{0.15,0.35,0.75},colorlinks=true,citecolor=[rgb]{0.675,0,0.2},urlcolor=[rgb]{0.15,0.35,0.65}}

\newcommand{\sumint}{\mathop{\sum\mkern-25mu\int}}

\def\beq{\begin{equation}}
	\def\eeq{\end{equation}}
\def\bsp#1\esp{\begin{split}#1\end{split}}
\def\d{{\rm d}}
\long\def\bea#1\eea{\begin{align}#1\end{align}}

\newcommand{\nn}{\nonumber}

\AtBeginDocument{
	\heavyrulewidth=.08em
	\lightrulewidth=.05em
	\cmidrulewidth=.03em
	\belowrulesep=.65ex
	\belowbottomsep=0pt
	\aboverulesep=.4ex
	\abovetopsep=0pt
	\cmidrulesep=\doublerulesep
	\cmidrulekern=.5em
	\defaultaddspace=.5em
}

\def\be{\begin{equation}}
	\def\ee{\end{equation}}

\usepackage{color}

\begin{document}

	\author{Wanchen Li}
	\email{wanchenli@fudan.edu.cn}
	\affiliation{Department of Physics, Center for Field Theory and Particle Physics, and Key Laboratory of Nuclear Physics and Ion-beam Application (MOE), Fudan University, Shanghai, 200433, China}
	
	\author{Xiaohui Liu}
	\email{xiliu@bnu.edu.cn}
	\affiliation{School of Physics and Astronomy, Beijing Normal University, and Key Laboratory of Multiscale Spin Physics (Beijing Normal University), Ministry of Education, Beijing 100875, China}
	\affiliation{Southern Center for Nuclear-Science Theory (SCNT), Institute of Modern Physics, Chinese Academy of Sciences, Huizhou 516000, Guangdong Province, China}

	\author{Ding Yu Shao}
	\email{dyshao@fudan.edu.cn}
	\affiliation{Department of Physics, Center for Field Theory and Particle Physics, and Key Laboratory of Nuclear Physics and Ion-beam Application (MOE), Fudan University, Shanghai, 200433, China}
	\affiliation{Shanghai Research Center for Theoretical Nuclear Physics, NSFC and Fudan University, Shanghai 200438, China}
	\affiliation{Center for High Energy Physics, Peking University, Beijing 100871, China}
	\affiliation{Southern Center for Nuclear-Science Theory (SCNT), Institute of Modern Physics, Chinese Academy of Sciences, Huizhou 516000, Guangdong Province, China}

	\title{Transverse Charge Distribution as a Probe of Nucleon Transversity}

	\begin{abstract}
		We introduce the transverse charge distribution as a spin-sensitive charge-flow probe for fragmentation and nucleon tomography. 
		By measuring the angular distribution of net electric charge around a fragmenting quark, this observable relies entirely on the tracking of charged-particle directions and charge signs, strictly bypassing the need for calorimetric energy measurements. 
		At leading twist, the distribution decomposes into an unpolarized charge monopole and a chiral-odd transverse charge dipole. 
		We derive the operator product expansion of these distributions onto charge-weighted collinear moments: a charge monopole and a charge dipole governed by the Collins effect and couples directly to transversity. 
		Applying this formalism to transversely polarized $p^\uparrow p$ collisions at RHIC, we show that charge weighting suppresses the unpolarized monopole background and causes the spin-dependent dipoles from oppositely charged hadrons to add coherently. 
		This coherence strongly enhances the resulting azimuthal asymmetries, establishing a theoretically clean and experimentally precise track-only avenue for extracting transversity.
	\end{abstract}

	\maketitle

	\paragraph*{Introduction.---}
	
	The transversity parton distribution function (PDF) ~\cite{Ralston:1979ys, Artru:1989zv, Jaffe:1991kp, Cortes:1991ja, Barone:2001sp} is an essential component of nucleon tomography, encoding the transverse polarization of quarks inside a transversely polarized nucleon. 
	Its first \(x\)-moment gives the nucleon tensor charge, a fundamental hadron-structure quantity, an important target for lattice QCD, and a sensitive probe of physics beyond the Standard Model~\cite{Jackson:1957zz, Lin:2017stx, Gao:2023ktu}.
	Due to its chiral-odd nature, the phenomenological extraction of transversity requires coupling it to a chiral-odd final-state matrix element.
	This is achieved primarily through the Collins fragmentation function (FF) in single-hadron production~\cite{Collins:1992kk, Collins:1993kq} or through dihadron interference fragmentation functions~\cite{Radici:2018iag, Cocuzza:2023oam, Cocuzza:2023vqs}.
	Together, these channels have enabled global extractions of transversity from semi-inclusive deep-inelastic scattering (SIDIS)~\cite{Bacchetta:2006tn, HERMES:2004mhh, HERMES:2010mmo, COMPASS:2012ozz, COMPASS:2014ysd} and hadron-in-jet measurements in polarized proton-proton collisions~\cite{Yuan:2007nd, Kang:2017btw, DAlesio:2017bvu, STAR:2017akg, STAR:2022hqg, DAlesio:2025jmr}.

	Despite recent progress in global extractions of transversity~\cite{Anselmino:2007fs, Anselmino:2015sxa, Kang:2015msa, Cammarota:2020qcw, Gamberg:2022kdb, Boglione:2024dal}, disentangling it from multidimensional hadron fragmentation information, including longitudinal energy fractions and transverse momenta, remains a substantial phenomenological challenge. 
	Moreover, such analyses require hadron identification and calibrated measurements of their momenta and energies, with uncertainties in the calorimeter energy-scale calibration imposing a major limitation on the achievable experimental precision.

	Angular correlators~\cite{Basham:1978bw, Basham:1978zq, Neill:2022lqx, Kang:2023gvg, Kang:2023big, Gao:2025evv, Bhattacharya:2025bqa, Cao:2025icu, Song:2025bdj, Gao:2025cwy, Kang:2026pro, Lee:2026hub, Chang:2013iba, Chang:2013rca, Chen:2020vvp, Li:2021zcf, Jaarsma:2022kdd, Chen:2022pdu, Jaarsma:2023ell, Lee:2023npz, Monni:2025zyv, Barata:2026eth} have recently emerged as probes of transversity through polarized fragmentation. 
	The one-point energy correlator (OPEC), measured over identified hadron species in transversely polarized $p^\uparrow p$ collisions~\cite{Gao:2025evv, STAR:2026epw}, has been proposed as a transversity-sensitive observable. 
	Its energy weighting requires per-particle energy reconstruction, whose precision is subject to calorimeter energy-scale uncertainties. The one-point charge correlator (OPCC)~\cite{Cao:2026fzq} instead takes advantage of the precisely determined directions and charge signs of charged tracks, requiring neither particle identification nor calorimetric information. 
	Existing OPCC studies use the net electric charge carried by particles at a given angle to probe unpolarized TMDs and the Sivers function. 
	Accessing transversity, however, requires a qualitatively different chiral-odd structure. Whether such a structure can arise in this track-based charge flow has remained a nontrivial theoretical question.
	
	In this Letter, we introduce the {\it transverse charge distribution}, a charge-flow observable that maps how hadronization redistributes electric charge in the transverse plane around a fragmenting parton, using only the charge sign and angle information of reconstructed tracks. 
	We find that transverse quark polarization does generate a leading-twist, chiral-odd charge dipole in this distribution. 
	The dipole measures the spin-dependent asymmetric displacement of positive and negative charge and provides a geometric imprint of the Collins effect~\cite{Collins:1992kk}. 
	The unpolarized component, in comparison, reduces to the OPCC studied previously. 
	It is rotationally symmetric, and its integral gives the net flow charge reaching the measured collinear region. This flow charge coincides with the fragmenting quark charge in perturbation theory, while nonperturbative hadronization could generate a deficit~\cite{Collins:2023cuo, Ke:2026unpub}.
	
	Phenomenologically, summing over reconstructed tracks with their charge signs suppresses the unpolarized component through cancellations between oppositely charged hadrons. 
	In the spin-dependent channel, however, the same charge signs compensate the reversal of the corresponding Collins responses, allowing the dipole contributions to add coherently and enhancing the azimuthal asymmetry. 
	This enhancement increases sensitivity to spin-dependent fragmentation and nucleon transversity in polarized measurements at RHIC and future electron-ion colliders in the U.S. and China~\cite{Accardi:2012qut, AbdulKhalek:2021gbh, Anderle:2021wcy}.

	\medskip
	\paragraph*{Transverse charge distribution.---} 
	We introduce the transverse charge distribution of a boosted quark as
	\bea
	&	{\cal J}^{q}_{\Gamma,\omega} ( {\vec{n}_T})  
	\nn \\ 
	= &
	\int  \d \xi \d^2{{x_T}} \,
	e^{i{\omega}\xi} \,
	{\rm Tr}\left[ \frac{\Gamma}{2}
	\langle 0|  \chi_n(\xi, {x_T} ) \widehat{\cal Q}(\hat{n}) {\bar \chi}_n(0) |0\rangle 
	\right] 
	\label{eq:TQD}\,,
	\eea 
	where \(\chi_n\) is the gauge-invariant collinear quark field along the light-cone direction $n$ in soft-collinear effective theory~\cite{Bauer:2000ew, Bauer:2000yr, Bauer:2001ct, Bauer:2001yt}.
	The Dirac matrix $\Gamma$ projects the spin state of the fragmenting quark, and $\widehat{\cal Q}(\hat n)$ is the charge-flow operator that measures the net electric charge emitted in the direction $\hat n = (\vec n_T,\cos\theta_n)$, where $\vec n_T= \sin\theta_n (\cos\phi_n,\sin\phi_n)$ is measured from the fragmenting quark. 
	For simplicity, we suppress the spin and color normalization factors, as well as possible regulator dependence. 
	
	\begin{figure}[t]
		\centering
		\includegraphics[width=0.85\linewidth]{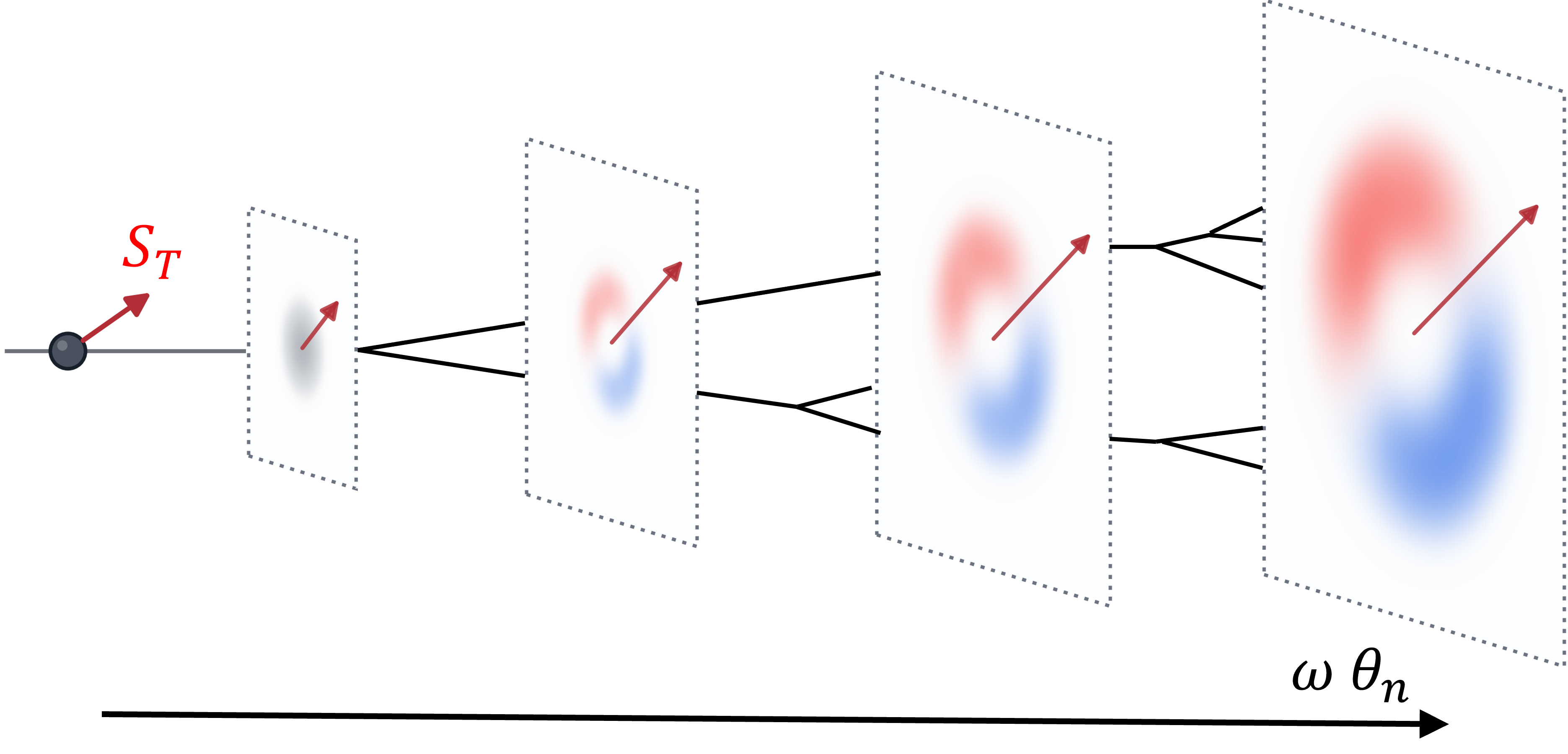}
		\caption{
			Schematic illustration of the transverse charge distribution generated in the fragmentation of a boosted transversely polarized quark. Distortions of the positive and negative charge distributions are shown in red and blue in the plane transverse to the fragmenting-quark direction. The dipole structure becomes more pronounced as the scale \(\omega\theta_n\) increases.}
		\label{fig:observable}
	\end{figure}
	
	The physical picture of the transverse charge distribution is illustrated in Fig.~\ref{fig:observable}. 
	A boosted quark with $\omega = p^+/2 $ fragments into hadrons, whose angular charge distribution is probed by a detector represented by the charge-flow operator~\cite{Hofman:2008ar, Monni:2025zyv, Cao:2026fzq} 
	\bea 
	{\widehat{\cal Q}}(\hat{n})
	=
	\lim_{r\to\infty} r^2
	\int_{-\infty}^{\infty} \d t\,
	\hat n_i J^i(t,r\hat{n}),
	\label{eq:charge-flow-op}
	\eea 
	where \(J^\mu\) is the electromagnetic current. 
	Acting on an asymptotic hadronic final state, the charge-flow operator admits the spectral representation
	\bea\label{eq:charge-flow-action} 
	\widehat{\cal Q}(\hat n) = 
	\sumint_X \, 
	\sum_{a \in X} Q_a \,
	\delta^{(2)}(\hat{n} -\hat{n}_{a})|X,\text{out}\rangle \langle X,\text{out}| \,, 
	\eea 
	which makes it explicit that this observable measures the charge-weighted angular distribution of final-state hadrons. 
	
	The transverse charge distribution is sensitive to the polarization of the initiating quark. 
	At leading twist, relevant quark spin projectors are $\Gamma = \gamma^+,\, \gamma^+ \gamma_5, \, i \sigma^{\alpha+}\gamma_5$, corresponding to an unpolarized, longitudinally polarized, and transversely polarized fragmenting quark, respectively. 
	Here, $\sigma^{\mu \nu} =\frac{i}{2}[\gamma^\mu,\gamma^\nu]$. 
	Because the final-state hadron polarization is unobserved by the charge-flow detector, parity conservation strictly forbids a non-zero charge distribution generated by a longitudinally polarized quark at leading twist~\cite{Bacchetta:2006tn}.
	Consequently, the leading non-zero structures are the unpolarized charge distribution and the transversely polarized, chiral-odd charge distribution
	\begin{align}
		\omega^{-2}	{\cal J}^{q}_{\gamma^+,\omega}({\vec{n}_T})
		&=
		J_D^q(\omega \theta_n),
		\label{eq:angular-monopole} \\
		\omega^{-2}	{\cal J}^{q,\alpha}_{i\sigma^{\alpha+}\gamma_5,\omega}({\vec{n}_T})
		&=
		\epsilon_T^{\alpha\beta}\, \omega \vec n_{T\beta}\,
		J_{1,\perp}^{q}(\omega  \theta_n),
		\label{eq:angular-dipole}
	\end{align}
	with  $\epsilon_T^{\alpha \beta} = \epsilon^{\alpha \beta \rho \sigma} \bar n_\rho n_\sigma/(n\cdot \bar n)$, where $\bar n \cdot n = 2$. 
	Here, we utilize the fact that in the collinear limit, the transverse distribution is longitudinal-boost invariant and therefore must be a function of $\omega  \vec n_T$~\cite{SupplementalMaterial}.
	
	It is useful to introduce the transverse charge distribution in impact parameter ($\vec b_T$) space. In the collinear limit, we have $\vec q_T = \omega \vec n_T$ with $q_T \equiv |\vec q_T|= \omega \theta_n$.
	Fourier transforming the spin-projected matrix element with respect to $\vec q_T$ yields~\cite{SupplementalMaterial}
	\bea 
	\tilde{{\cal J}}^q_\Gamma (\vec b_T)  = \int \d^2 \vec q_T\,  e^{i\vec q_T \cdot \vec b_T} {\cal J}_\Gamma^q(\vec q_T) \,.
	\eea 
	Writing the leading spin structures explicitly gives  
	\begin{align}
		\tilde{J}^{q}_D(b_T ) &=2\pi \int {\d q_T}\, q_T\,J_0(q_T b_T)\,{J}_D^{q}( q_T), \\
		b_T\,\tilde{J}^q_{1,\perp}(b_T ) & = 2\pi  \int {\d q_T}\,q_T^2 \, J_1(q_T b_T)\,{J}^q_{1,\perp}( q_T ) \,, 
		\label{eq:Fourier_transform}
	\end{align}
	where $b_T = |\vec b_T|$ and $J_0$, $J_1$ are Bessel functions.
	
	\medskip
	
	\paragraph*{Operator Product Expansion.---}
	When $b_T\Lambda_{\rm QCD}\ll1$, the transverse charge distributions admit an OPE onto collinear nonperturbative matrix elements. The OPE can be derived directly from the charge-flow operator definition, making its transverse multipole interpretation manifest. It can also be obtained by relating the transverse charge distributions to charge-weighted moments of standard TMD FFs. The latter relation is not required for defining the observable, but provides a convenient connection to the standard TMD factorization and evolution framework used below. Details of both derivations are given in the Supplemental Material~\cite{SupplementalMaterial}. The resulting OPEs are
	\begin{align}
		\tilde J^{q}_{D}(b_T,\mu,\zeta) &= \sum_j \tilde C^D_{j/q} (b_T,\mu,\zeta) \,
		Q_{{\rm flow}\,, j} \,,  
		\label{eq:OPE-Unpolarized}\\
		\tilde J^{q}_{1,\perp}(b_T,\mu,\zeta) & = \sum_j   \tilde  C^{\rm Collins}_{j/ q}(b_T,\mu,\zeta)\, 
		{\cal D}^\perp_j(\mu)\,.
		\label{eq:OPE-Collins}
	\end{align}
	Here $\mu$ and $\zeta$ are the renormalization and Collins-Soper scales, respectively, and the sum runs over quark flavors $j$.
	In the unpolarized channel, the nonperturbative matrix element reduces to the flow charge $Q_{{\rm flow}\,, j}$ measured by a detector and is scale independent. In perturbation theory, $Q_{{\rm flow}\,, j}$ coincides with the quark charge $Q_j$ \cite{Monni:2025zyv}.
	Hadronization could modify this scenario~\cite{Collins:2023cuo, Ke:2026unpub}. 
	In the transverse-spin channel, the nonperturbative coefficient ${\cal D}_j^\perp(\mu)$ can be related to the charge-weighted moment of the Collins function $H_{1h/j}^{\perp(1)}(z,\mu)$ through
	\begin{align}    \label{eq:Collins_dipole}
		{\cal D}_j^\perp(\mu)
		&= \sum_h\int \d z \,Q_h \,M_h \,H^{\perp(1)}_{1h/j}(z,\,\mu)  \\ 
		&    = \sum_{\bar h}\int \d z\, (-Q_{\bar h})\, M_{\bar h}H_{1{\bar h}/{\bar j}}^{\perp(1)}(z,\mu)
		= - 
		{\cal D}_{\bar j}^\perp(\mu) \, . \nn 
	\end{align}
	Here $M_h$ balances the mass dimensions in the Trento convention~\cite{Bacchetta:2004jz}, and the last equality follows from charge conjugation. Eq.~\eqref{eq:Collins_dipole} identifies ${\cal D}_j^\perp$ as the intrinsic transverse charge dipole dynamically generated by the Collins effect. We note that the $z$ integral arises only if we choose to relate the charge-flow operator to the TMD FFs through its hadronic spectral decomposition. The charge-flow detector performs the inclusive sum directly, without measuring $z$ or reconstructing a $z$ integral. 

	The relation to the standard TMD FFs further indicates that the short-distance coefficients in Eqs.~\eqref{eq:OPE-Unpolarized} and~\eqref{eq:OPE-Collins} can be obtained from the usual TMD matching coefficients
	\begin{align}
		&\tilde C_{j/q}^F(b_T,\mu,\zeta) = \notag \\
		&\quad \int_0^1 \d z \left[  C_{j/q}^F(z,b_T,\mu,\zeta)
		- C_{{\bar j}/q}^F(z,b_T,\mu,\zeta)\right],
	\end{align}
	with $F\in\{D,\, {\rm Collins}\}$. 
	The unpolarized matching coefficients $\tilde{C}^D$ are known to N$^3$LO \cite{Monni:2025zyv,vita_2025_16658462,Luo:2020epw}, while the transversely polarized coefficients $ C^{\rm Collins}$ are known up to NLO \cite{Koike:2006fn, Yuan:2009dw, Bacchetta:2013pqa, Echevarria:2014rua, Kang:2015msa}. 
	At leading order, both matching coefficients reduce to $\delta_{jq}$. 
	We note that a complete twist-3 collinear treatment also contains contributions from quark-gluon-quark correlators~\cite{Yuan:2009dw,Kang:2010xv,Metz:2012ct,Scantamburlo:2026hcp}. Their phenomenological impact remains largely unexplored and is deferred
	to future work.

	\medskip
	
\begin{figure}[t]
		\centering
		\includegraphics[width=0.75\linewidth]{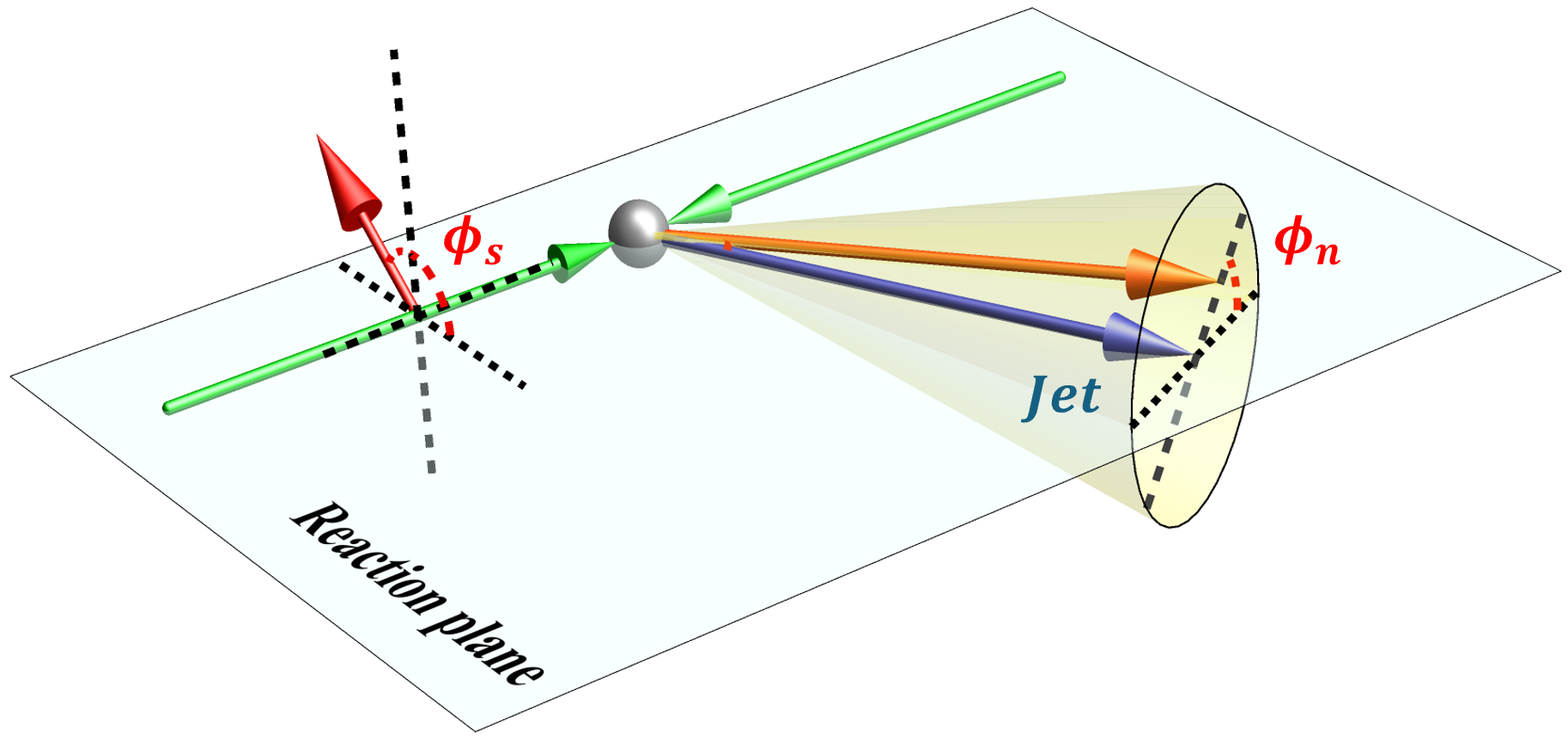}
		\caption{Geometry of transverse charge-flow measurements inside a jet in \(p^\uparrow p\) collisions. The azimuthal angles \(\phi_s\) and \(\phi_n\) are measured with respect to the reaction plane defined by the beam and jet axes.}
		\label{fig:pp_collision}
	\end{figure}

	\paragraph*{RHIC phenomenology.---}
	Now we demonstrate how the transverse charge distribution probes the incoming proton transversity. 
	We consider measurements of charge flow inside jets in transversely polarized proton-proton collisions,
	\begin{align}
		\frac{\d\Sigma_Q(\hat n)}
		{\d\eta\,\d p_T}   
		=
		\sum_{h\in J}
		\int \d^2\Omega_h\,
		Q_h\,
		\delta^{(2)}(\hat n - \hat n_h)
		\frac{\d\sigma}
		{
			\d^2\Omega_h\,\d\eta\,\d p_T},
		\label{eq:charge-phen-def}
	\end{align}
	Here the sum runs over charged hadrons in the jet. 
	The jet is characterized by its transverse momentum \(p_T\) and rapidity \(\eta\). 
	As illustrated in \cref{fig:pp_collision}, the incoming beams and the jet axis define the reaction plane, relative to which the proton-spin azimuth $\phi_s$ and the charge-flow azimuth $\phi_n$ are measured.
	
	For small $\theta_n$, the charge flow cross section takes the azimuthal form
	\begin{equation}
		\frac{d\Sigma_Q(\hat n)}
		{d\eta\,dp_T}
		=
		Z_{UU}^Q
		+
		\sin(\phi_s-\phi_n)\,Z_{UT}^Q.
		\label{eq:charge-azimuthal-decomp}
	\end{equation}
	Given the connection between the transverse charge distribution and the TMD FFs, we can immediately show that the unpolarized ($Z_{UU}^Q$) and spin-dependent ($Z_{UT}^Q$) structure functions factorize in the same way as those for hadron-in-jet process with the FFs replaced by the transverse charge distribution, which gives 
	\begin{align}
		Z_{UU}^Q
		=&\;
		\frac{\alpha_s^2}{s}\,
		p_T^2\theta_n
		\sum_{a,b,c}
		\int \frac{dx_a}{x_a}\,
		f_{a/A}(x_a,\mu)
		\int \frac{dx_b}{x_b}\,
		f_{b/B}(x_b,\mu)
		\nonumber\\
		&\hspace{-1cm}\times
		H_{ab\to c}^{U}(\hat s,\hat t,\hat u,\mu)\,
		J_D^c({p_T}\theta_n)\,
		\delta(\hat s+\hat t+\hat u),
		\label{eq:ZUU-charge}
		\\
		Z_{UT}^Q
		=&\;
		\frac{\alpha_s^2}{s}\,
		p_T^2\theta_n
		\sum_{a,b,c}
		\int \frac{dx_a}{x_a}\,
		h_1^{a}(x_a,\mu)
		\int \frac{dx_b}{x_b}\,
		f_{b/B}(x_b,\mu)
		\nonumber\\
		&\hspace{-1cm}\times
		H_{ab\to c}^{\rm Collins}(\hat s,\hat t,\hat u,\mu)\,
		p_T\theta_n\,J_{1,\perp}^c({p_T} \theta_n)\,
		\delta(\hat s+\hat t+\hat u). 
		\label{eq:ZUT-charge}
	\end{align}
	Here, the soft function has been absorbed into the transverse charge distributions $J_{D}^c$ and $J_{1,\perp}^c$. 
	$\hat s$, $\hat t$,  and  $\hat u$ are the partonic Mandelstam variables. \(f\) is the unpolarized collinear PDF, and \(h_1^a\) is the transversity distribution of the polarized proton.  
	The hard functions \(H_{ab\to c}^{U}\) and \(H_{ab\to c}^{\rm Collins}\) describe the unpolarized hard scattering and transverse-spin transfer to the parton $c$ that initiates the observed jet, respectively~\cite{Owens:1986mp, Yuan:2007nd, Kang:2010zzb, Kang:2017btw}. 
	Jet algorithm dependence can be incorporated within the fragmenting jet function framework ~\cite{Kang:2017glf, Kang:2023elg, Gao:2025evv}. 

	For $\theta_n\gg\Lambda_{\rm QCD}/\omega$, the transverse charge distributions can be evaluated using TMD evolution
	\begin{align}
		J_D^q(p_T \theta_n)
		=&\,
		\int_0^\infty
		\frac{\d b_T\,b_T}{2\pi}\,
		J_0(b_T  p_T \theta_n)\,e^{-\frac{1}{2}S_{\rm pert}^{D}(\omega,b_T)}
		\nonumber\\
		&\times \sum_j \tilde C^D_{j / q} (b_T,\mu_b) \, 
		Q_{{\rm flow}\,, j},
		\label{eq:JD-theta}
		\\
		\hspace{-0.2cm} p_T \theta_n\,J_{1,\perp}^q({p_T}\theta_n)
		=&\,
		\int_0^\infty
		\frac{\d b_T\,b_T^2}{2\pi}\,
		J_1(b_T p_T \theta_n)\, e^{-\frac{1}{2}S_{\rm pert}^{\rm Collins}(\omega,b_T)}
		\nonumber\\
		&\times  
		\sum_j \tilde  C^{\rm Collins}_{j/ q}(b_T,\mu_b)\,{\cal D}_j^\perp({\mu_b}),
		\label{eq:Jperp-theta}
	\end{align}
	where $\omega \sim p_T$ is the hard scale and  \(\mu_b=2e^{-\gamma_E}/b_T\). 
	We adopt the \(b^*\)-prescription to avoid the Landau pole with $b^*={b_T}/{\sqrt{1+b_T^2/b_{\rm max}^2}}$ and $b_{\rm max} = 1.5~\text{GeV}^{-1}$~\cite{Collins:1984kg, Qiu:2000ga, Landry:2002ix}. 
	The Sudakov factors~\cite{Sun:2014dqm, Kang:2015msa} contain the TMD evolution effects.

	\begin{figure}[t]
		\centering
		\includegraphics[width=0.9\linewidth]{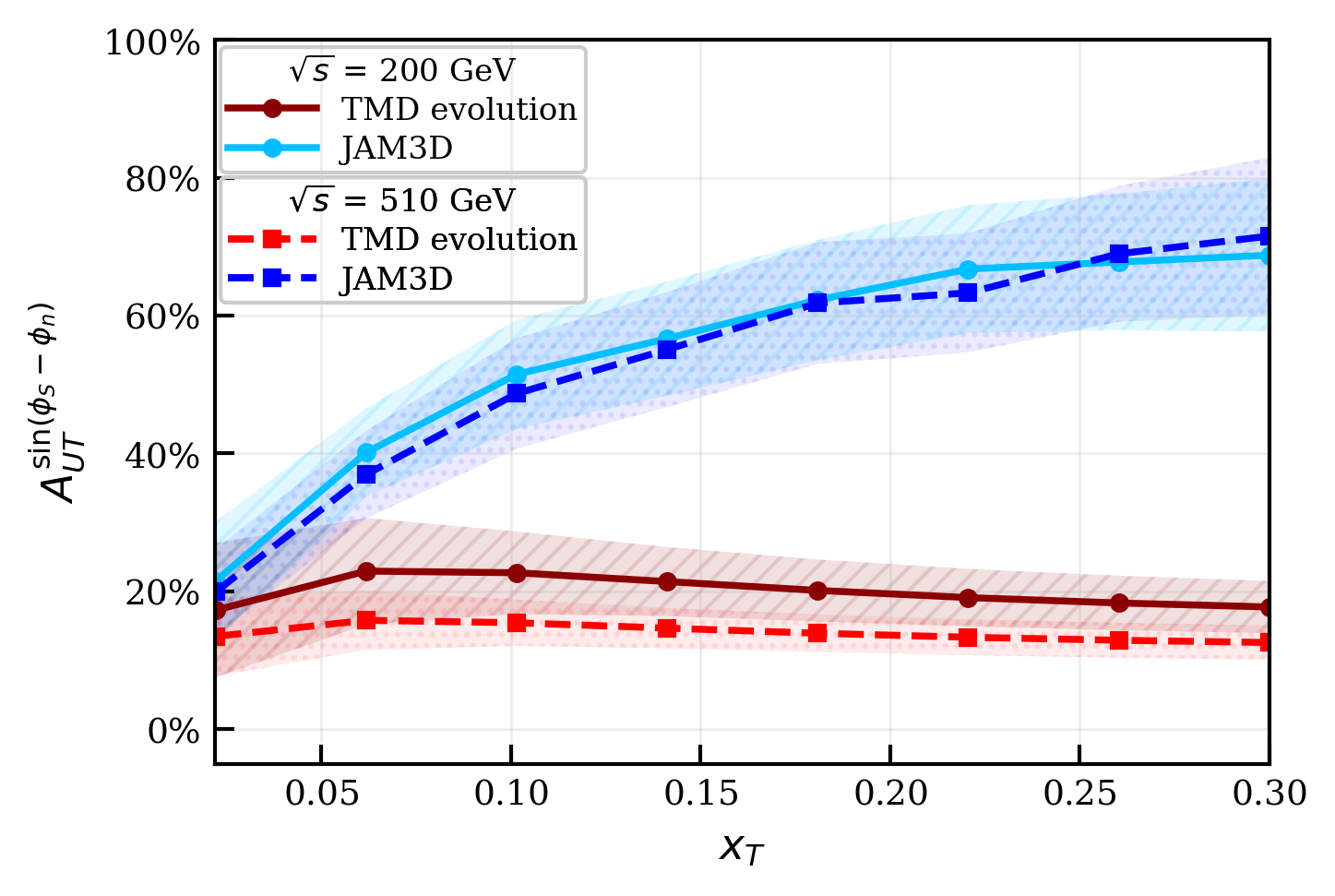}
		\caption{The transverse charge distribution SSA $A_{UT}^{\sin(\phi_s - \phi_n)}$ for charged pions as a function of $x_T = 2p_T/\sqrt{s}$ at center-of-mass energies $\sqrt{s} = 200$ GeV (solid curves) and $510$ GeV (dashed curves), integrated over \(0<\theta_n<0.4\), in the JAM3D and TMD-evolution frameworks. Uncertainties are propagated from parametrizations of transversity PDFs and Collins functions.}
		\label{fig:SSA_pT}
	\end{figure}
	
	For the phenomenological study, since the charge dipole ${\cal D}_j^\perp$ has not been measured, we estimate its size from the charged-pion Collins functions available from global analyses.   
	We present predictions using two approaches. 
	The first uses the TMD-evolution framework in \cref{eq:JD-theta,eq:Jperp-theta} with NLL TMD resummation and LO OPE matching coefficients. The transversity distribution, Collins function, and nonperturbative Sudakov factors are taken from the global analysis Ref.~\cite{Kang:2015msa}. 
	The second uses the JAM3D-22 global analysis~\cite{Gamberg:2022kdb}, where the transverse-momentum dependence is modeled by Gaussian profiles and the collinear components are only evolved by the DGLAP equations. 
	The comparison between these two approaches provides a useful estimate of the present sensitivity to TMD evolution and to the modeling of transverse momentum dependence.

	In \cref{fig:SSA_pT}, we show the transverse charge distribution single spin asymmetry (SSA), $A_{UT}^{\sin(\phi_s - \phi_n)} = {Z_{UT}^Q}/{Z_{UU}^Q}$, as a function of $x_T=2p_T/\sqrt{s}$. 
	We select jets with pseudorapidity $0<\eta<0.9$ with respect to the polarized proton beam.
	For this \(p_T\)-differential evaluation, the structure functions are integrated over \(0<\theta_n<0.4\).
	The two collision energies, $\sqrt{s}=200$ and $510~\mathrm{GeV}$, lead to very similar asymmetries when compared at fixed $x_T$, consistent with the energy independence observed in recent STAR measurements of TMD spin asymmetries~\cite{STAR:2025xyp}.
	
	\begin{figure}[t]
		\centering
		\includegraphics[width=0.9\linewidth]{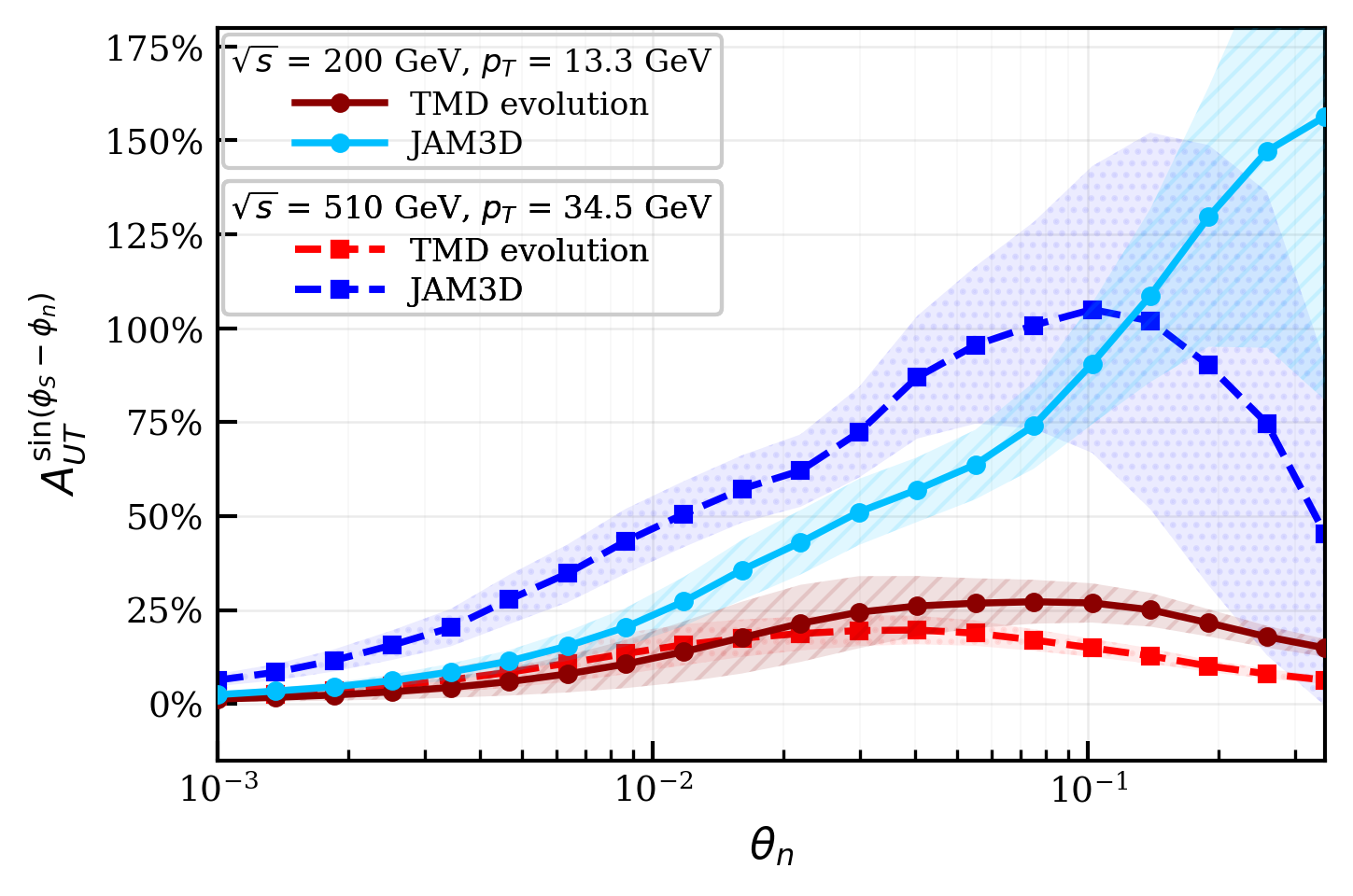}
		\caption{The transverse charge distribution SSA $A_{UT}^{\sin(\phi_s - \phi_n)}$  for charged pions as a function of $\theta_n$ at $\sqrt{s} = 200\, \text{GeV}$, $p_T = 13.3\, \text{GeV}$ (solid curves) and $\sqrt{s} = 510\, \text{GeV}$, $p_T = 34.5\, \text{GeV}$ (dashed curves), evaluated in the JAM3D and TMD-evolution frameworks. Uncertainties are propagated from parametrizations of transversity PDFs and Collins FFs.}
		\label{fig:SSA_theta}
	\end{figure}

	In \cref{fig:SSA_theta}, we show the \(\theta_n\) dependence of the SSA for two representative STAR kinematics, \(\sqrt{s}=510~\mathrm{GeV}\) with \(p_T=34.5~\mathrm{GeV}\) and \(\sqrt{s}=200~\mathrm{GeV}\) with \(p_T=13.3~\mathrm{GeV}\), which probe comparable values of \(x_T\).
	Compared with the corresponding non-charge-weighted observables, the transverse charge asymmetry is strongly enhanced, reaching \(20\%\)--\(25\%\) in the TMD-evolution framework and exceeding unity in magnitude at larger \(\theta_n\) in JAM3D.
	We note that values above unity do not imply a positivity violation, because \(Z_{UU}^Q\) is a charge-weighted monopole moment rather than a positive-definite yield and is strongly suppressed by cancellations between oppositely charged hadrons.
	At the same time, \(Z_{UT}^Q\) is approximately twice its non-charge-weighted counterpart because the charge dipoles of oppositely charged pions add coherently after charge weighting.
	
	This coherent enhancement in the Collins channel also indicates that the transverse charge distribution need not suffer a loss of relative statistical precision compared with the separate positively and negatively charged hadron distributions.
	We further illustrate this point in the Supplemental Material \cite{SupplementalMaterial} using transversely polarized SIDIS events generated by PYTHIA 8.3~\cite{Bierlich:2022pfr} with StringSpinner \cite{Kerbizi:2021pzn, Kerbizi:2022hhs, Kerbizi:2023cde, Kerbizi:2026iws} at COMPASS and EIC~\cite{Accardi:2012qut, AbdulKhalek:2021gbh} kinematics.
	
	The enhancement of the JAM3D predictions in the large-$\theta_n$ region is mainly driven by the transverse-momentum profiles. 
	Although the unfavored collinear component is smaller, its unpolarized TMD FF has a slightly broader Gaussian width, \(\langle P_\perp^2\rangle_D^{\rm unfav}\simeq 0.144~\mathrm{GeV}^2\), than the favored one, \(\langle P_\perp^2\rangle_D^{\rm fav}\simeq 0.127~\mathrm{GeV}^2\)~\cite{Gamberg:2022kdb}, making it relatively more important at large \(\theta_n\).
	This enhances the cancellation in \(Z_{UU}^Q\) and amplifies the charge-weighted asymmetry.
	This sensitivity suggests that transverse charge asymmetries can provide strong constraints on the flavor and transverse momentum dependence of TMD FFs, and may help discriminate between descriptions with and without TMD evolution.
	Related uses of charge weighting for flavor discrimination have also been explored with jet charge~\cite{Waalewijn:2012sv,Krohn:2012fg,Kang:2020fka}.

	\medskip
	
	\paragraph*{Conclusions.---}
	We introduced the transverse charge distribution, which contains a leading-twist, chiral-odd dipole generated by a transversely polarized quark. The dipole measures the spin-dependent transverse displacement of positive and negative charge during hadronization and provides a geometric representation of the Collins effect. At short transverse distances, the rotationally symmetric monopole matches onto the scale independent flow charge, while the dipole matches onto the charge-weighted Collins moment ${\cal D}_j^\perp $. 
	This chiral-odd dipole supplies the final-state matrix element that couples to nucleon transversity in polarized scattering.
	
	Measuring the transverse charge distribution in transversely polarized proton-proton collisions produces a distinctive enhancement of the spin modulation. Oppositely charged hadrons cancel in the signed unpolarized monopole. In the polarized channel, the electric charges compensate the reversal of the corresponding Collins responses, so the spin-dependent dipoles add coherently. The simultaneous suppression of $Z_{UU}^Q$ and enhancement of $Z_{UT}^Q$ yield asymmetries of $20\%$--$25\%$ in our TMD-evolution estimates. The comparison with JAM3D further demonstrates sensitivity to the transverse-momentum structure of polarized fragmentation.
	
	The charge-flow measurement uses only track directions and charge signs, without identified-hadron tagging or per-hadron energy or energy-fraction measurements. It also reduces the fragmentation-specific nonperturbative information from a multidimensional Collins TMD distribution to the universal charge dipole ${\cal D}_j^\perp $, a single number for each parton flavor. The present study estimates this number from charged-pion Collins fits, while future transverse-charge measurements can determine it directly. The transverse charge distribution thus provides a species-inclusive, track-based channel for jointly constraining the charge dipole and nucleon transversity at RHIC and future electron-ion colliders.
	
	\medskip
	
	\paragraph*{Acknowledgments.---}
	We thank Weiyao Ke and Haitao Li for helpful discussions. 
	W.~L. and D.~Y.~S. are supported by the National Natural Science Foundation of China under Grant No.~12275052, No.~12147101, No.~12547102, and the Innovation Program for Quantum Science and Technology under grant No. 2024ZD0300101.
	W.~L. is also supported by the China Postdoctoral Science Foundation under grant No.~2025M783369.
	X.~L. is supported by the National Natural Science Foundation of China under Grant No.~12547109 and Fundamental Research Funds for the Central Universities, Beijing Normal University.

	\bibliographystyle{apsrev4-1}
	\bibliography{refs.bib}

	\clearpage
	\appendix
	\onecolumngrid
	\setcounter{equation}{0}
	\renewcommand{\theequation}{S-\arabic{equation}}
	\setcounter{figure}{0}
	\renewcommand{\thefigure}{S-\arabic{figure}}
	\renewcommand{\theHfigure}{S-\arabic{figure}}
	\allowdisplaybreaks
	\section*{Supplemental Material}

	\subsection*{Variable transformation}
	We make the mapping from the detector direction $\hat n$ to the transverse variable $q_T$ explicit. 
	We stereographically project the detector direction $\hat n$ onto the transverse plane and define the projective transverse coordinate 
	\bea 
	\vec \rho_T (\hat n) \equiv \frac{2 \vec n_T}{1+\cos\theta}
	= 2 \tan \left(\frac{\theta}{2}\right) (\cos\phi,\sin\phi) \,.
	\eea 
	We then introduce the transverse vector $\vec q_T$ 
	\bea 
	\vec q_T \equiv \omega \vec \rho_T (\hat n) \,. 
	\eea 
	For $\theta \in [0,\pi)$, 
	the stereographic map covers the entire transverse plane, with
	$|\rho_T|$ and $|\vec q_T|$ defined on $[0,\infty)$. The point $\theta = \pi$, corresponding to the direction $\bar n$, is mapped to transverse infinity.
	
	At leading power in the collinear expansion, $\theta \ll 1$, the projective coordinate reduces to the transverse angular vector 
	\bea 
	\vec \rho_T (\hat n)  \approx \vec n_T   \,, \qquad 
	\vec q_T \approx  \omega \vec n_T   \,, 
	\eea 
	up to relative power corrections of order ${\cal O}(\theta^2)$. This angular parametrization agrees with main text $\vec n_T= \sin\theta_n (\cos\phi_n,\sin\phi_n)$ in the collinear limit. 
	
	\subsection*{Boost Invariance Constraint}~\label{app:rpi}
	
	For the transverse charge distribution in eq.~(1), we consider a boost along the fragmenting quark direction $n$. 
	Under this transformation, the quark energy $\omega$ scales as
	\bea 
	\omega \to \omega' = e^Y \omega \,, 
	\eea 
	where $Y$ is the boost rapidity. 
	The corresponding transverse angle transforms inversely according to
	\bea 
	\vec n_T \to \vec n_T' = e^{-Y} \vec n_T \,,
	\eea 
	and $\d^2  \vec n_T' = e^{-2Y} \d^2\vec n_T$.
	Since the fragmentation matrix element cannot depend on the arbitrary longitudinal frame used to describe the same collinear process, the transverse charge distribution must satisfy
	\begin{equation}
		{\cal J}_{\Gamma,\omega}^q(\vec n_T) 
		=e^{-2Y} {\cal J}_{\Gamma,\omega e^Y}^q(e^{-Y}\vec n_T) \,,
	\end{equation}
	in the collinear limit, $|\vec n_T| \ll 1$ for any boost rapidity $Y$. 
	For an infinitesimal boost, this leads to 
	\bea 
	\left( \omega \, \partial_\omega  - \vec n_T \cdot \vec{\partial}_{\vec n_T} -2
	\right) {\cal J}_{\Gamma,\omega}^q(\vec n_T) = 0 \,,
	\eea 
	whose general solution takes the form
	\bea 
	\omega^{-2}
	{\cal J}_{\Gamma,\omega}^q(\vec n_T) =  {\cal J}_\Gamma^q(\omega \, \vec n_T),
	\eea 
	where we can define $ {\cal J}_\Gamma^q(\vec q_T ) \equiv \omega^{-2}
	{\cal J}_{\Gamma,\omega}^q(\vec n_T)$ with $\vec q_T = \omega \, \vec n_T$.
	Consequently, the unpolarized and transversely polarized charge distributions can be parameterized as 
	\begin{align}
		{\cal J}^{q}_{\gamma^+}(\vec q_T )
		&=
		J_D^q(q_T ),
		\\
		{\cal J}^{q,\alpha}_{i\sigma^{\alpha+}\gamma_5}(\vec q_T )
		&=
		\epsilon_T^{\alpha\beta}\, \frac{q_{T,\beta}}{\Lambda_{\rm QCD}}\,
		{J}_{1,\perp}^{\prime q}(q_T).
	\end{align}
	We note that $J_{1,\perp}^{q}(q_T)$ in eq.~(5) has mass dimension $(-3)$, satisfying the relation $ J_{1,\perp}^{q}(q_T) = \frac{1}{\Lambda_{\rm QCD}}{J}_{1,\perp}^{\prime q}(q_T) $.

	\subsection*{Operator Product Expansions of the Transverse Charge Distribution}~\label{app:tmd}
	We derive the operator product expansions (OPEs) of the transverse charge distribution presented in eqs.~(9)--(12) of the main text. 
	This derivation is carried out via two complementary approaches: first, by establishing a relation between the transverse charge distribution and standard TMD FFs; second, by directly performing a transverse multipole expansion on the charge-flow operator definition given in eq.~(1) of the main text.

	\subsubsection{Relation to TMDs}
	
	The charge-flow matrix element can be mapped onto standard TMD FFs by resolving the inclusive charge measurement into identified hadrons.
	In the parton frame, the TMD FF $d_{\Gamma,h/q}(z,\vec p_{hT})$ shares the core correlator structure of eq.~(1)~\cite{Collins:2011zzd}, differing only by the replacement of the continuous charge-flow operator $\widehat{\cal Q}$ with the discrete hadron-tagging measurement operator ${\cal M}_h$:
	\bea
	& d_{\Gamma,h/q}(z, \vec p_{hT})
	=
	\int {\d \xi \d^2{{x_T}}}\,
	e^{i{\omega}\xi} \, {\rm Tr}\left[ \frac{\Gamma}{2} 
	\langle 0|  \chi_n(\xi, {x_T} ) \,{\cal M}_h\, {\bar \chi}_n(0) |0\rangle  \right] ,\label{eq:fragmentation}
	\eea 
	where 
	\begin{align}\label{eq:tmd-measure}
		{\cal M}_h(z, \vec p_{hT}) = 
		\sumint_X \, 
		\sum_{a\in X} \delta_{ah} \,
		\delta\left(z-\frac{p_a^+}{2\omega}\right) 
		\,\delta^{(2)}(\vec p_{hT}-\vec p_{aT}) |X,\text{out}\rangle \langle X,\text{out}|\,,
	\end{align}
	which measures the momentum fraction $z$ and the transverse momentum $\vec p_{hT}$ of the identified hadron $h$ in the collinear sector.  
	
	Comparing this definition with eqs.~(1) and~(3), we find that the two observables share the same quark-field correlator but differ in the measurement operators inserted between the quark fields. 
	The fragmentation process itself is encoded in the collinear correlator initiated by the parent quark, whereas ${\cal M}_h$ specifies how that process is probed by tagging an identified hadron. 
	In this sense, hadron identification in fragmentation functions is therefore one possible measurement of fragmentation, rather than a unique choice. 
	It could be replaced by more inclusive probes whose theoretical or experimental properties are advantageous for a given application. 
	Energy-flow correlators provide one example of this strategy, and the charge-flow measurement considered here provides another.
	
	Performing a boost from the parton frame to the hadron frame maps the angular variable $\vec q_T\equiv\omega\vec n_T$ directly to the hadron transverse momentum, $\vec p_{hT}=z\,\vec q_T$~\cite{Collins:2011zzd}. 
	Integrating over phase space then relates the transverse charge distribution to the charge-weighted sum over regular TMD FFs:
	\begin{align}
		{\cal J}^{q}_\Gamma(\omega \vec{n}_T)  
		& = 
		\sum_h Q_h \int \d z\, z^2\, d_{\Gamma,h/q}(z, z  \omega \vec n_T) \,.
		\label{eq:J_to_TMD}
	\end{align} 
	Here, the summation is over the hadrons $h$ within the collinear sector. 
	Fourier transformation with respect to $\vec q_T$ yields the corresponding relation in $\vec b$ space,  
	\begin{align}
		\tilde{{\cal J}}^{q}_\Gamma( \vec b_T)  
		& =  
		\sum_h Q_h \int \d z\, \, \tilde{d}_{\Gamma,h/q}(z, \vec b_T) \,. 
		\label{eq:J_to_TMD_b}
	\end{align} 
	With the relation to TMD FFs established, it is straightforward to derive the OPEs of the transverse charge distributions in Eqs.~(9)--(12) in the main text.

	\subsubsection{Direct OPE from the Charge-Flow Definition}
	
	The small-$b_T$ OPEs can alternatively be derived directly from the charge-flow definition in eq.~(1), completely bypassing the intermediate TMD FF representation.
	In this formulation, the small-$b_T$ expansion corresponds to a transverse multipole expansion of the charge-flow distribution. The leading monopole measures the net charge reaching the detector, and the transverse charge dipole is associated with the Collins moment. 
	We begin by introducing the spectral decomposition of the charge flow detector 
	\bea
	\widehat{\cal Q}(\vec q_T ) = 
	\sumint_X \, 
	\sum_{a \in X} Q_a \,
	\delta^{(2)}(\vec q_T -\vec p_{aT}/z_a)|X,\text{out}\rangle \langle X,\text{out}| \,, 
	\eea 
	where we have rewritten the charge flow in terms of the transverse momentum variable $\vec q_T = \omega \vec n_T$, up to a trivial normalization factor of $\omega^2$. 
	Although a spectral decomposition is not required for the direct OPE, we introduce it here to make the hadronic interpretation of the transverse multipoles manifest.

	The Fourier transformation of the transverse charge distribution with respect to $\vec q_T$ acts only on the operator $\widehat{\cal Q}(\vec q_T) $. Its Fourier transform is
	\bea 
	\widehat{{\cal Q}}(\vec b_T)
	= \int d^2 \vec q_T e^{i\vec q_T \cdot \vec b_T} \widehat{{\cal Q}}(\vec q_T) 
	= \int d^2 \vec q_T 
	\left[
	1 + i \vec q_T \cdot \vec b_T 
	+ \dots 
	\right] \widehat{{\cal Q}}(\vec q_T) \,,
	\eea \
	where the $\vec b_T$ expansion has the interpretation of a transverse multipole expansion of the charge-flow distribution. 
	The leading term measures the monopole, namely the net charge that reaches the detector whereas the term linear in $b_T$ measures the dipole moment of the charge distribution in transverse recoil space. 
	In terms of the spectral decomposition, they read    
	\bea 
	\int d^2 \vec q_T  \,
	\hat{{\cal Q}}(\vec q_T) & = \hat{Q}_{\rm flow} 
	= 
	\sumint_X \, 
	\sum_{a \in X} Q_a \,
	|X,\text{out}\rangle \langle X,\text{out}| 
	\,,  \nn \\ 
	i \vec b_T  \cdot \left[  
	\int d^2 \vec q_T \,
	\vec q_T  \,
	\hat{{\cal Q}}(\vec q_T)
	\right] &
	= i \vec b_T \cdot {\hat{\vec {\cal D}}}_T
	= i \vec b_T \cdot 
	\sumint_X \, 
	\sum_{a \in X} 
	Q_a \, 
	\frac{ \vec p_{aT} }{z_a}
	|X,\text{out}\rangle \langle X,\text{out}|  \,. 
	\eea 
	We defer a detailed discussion of $\hat{\cal Q}_{\rm flow}$ to future work~\cite{Ke:2026unpub}. 
	This form manifests that the dipole moment ${\hat{\vec {\cal D}}}_T$ receives contributions from the first $\vec p_{aT}/z_a$ moment of the asymptotic hadrons. 
	The first nontrivial spin information about the transverse charge distribution is thus contained in the dipole operator ${\hat{\vec {\cal D}}}_T$.
	Physically, this operator measures whether positive and negative charge are displaced asymmetrically in transverse momentum space during hadronization.
	Restricting our attention to this dipole term, we insert the charge-flow operator back into eq.~(1) to extract the non-vanishing dipole contribution
	\bea
	ib_{T\alpha} {\cal M}^{\rho\alpha}
	=
	i   b_{T\alpha}  
	\int  \d \xi \d^2 \vec x_T \,
	e^{i{\omega}\xi} \,
	{\rm Tr}\left[ \frac{i\sigma^{\rho+}\gamma_5}{2}
	\langle 0|  \chi_n(\xi, {x_T} ) \, 
	{\widehat{{\cal D}}}^\alpha_T   \, 
	{\bar \chi}_n(0) |0\rangle 
	\right]  \,, 
	\eea 
	which describes a fragmentation process initiated by a transversely polarized quark that generates a nonvanishing transverse charge dipole.    
	Here, we define 
	\bea 
	{\cal M}^{\rho\alpha}
	& \equiv 
	\int  \d \xi \d^2 \vec x_T  \,
	e^{i{\omega}\xi} \,
	{\rm Tr}\left[ \frac{i\sigma^{\rho+}\gamma_5}{2}
	\langle 0|  \chi_n(\xi, {x_T} )  
	\sumint_X \, 
	\sum_{a \in X} 
	Q_a \, 
	\frac{ p_{aT}^\alpha  }{z_a}
	|X,\text{out}\rangle \langle X,\text{out}|    
	{\bar \chi}_n(0) |0\rangle 
	\right]  \\ 
	& = \sum_h Q_h \,
	\int d\Phi_h 
	\frac{ p_{hT}^\alpha }{z_h}
	\, 
	\int  \d \xi \d^2 \vec x_T  \,
	e^{i{\omega}\xi} \,
	{\rm Tr}\left[ \frac{i\sigma^{\rho+}\gamma_5}{2}
	\langle 0|  \chi_n(\xi, {x_T} )  
	\sumint_{X'} \,  
	|h(\Phi_h), X',\text{out}\rangle \langle h(\Phi_h) ,X',\text{out}|    
	{\bar \chi}_n(0) |0\rangle 
	\right]   \,, \nn 
	\eea 
	where $\int d\Phi_h = \int \frac{dp_h^+ d^2\vec p_{hT}}{2(2\pi)^3 p_h^+} = \int \frac{dz_h d^2\vec p_{hT}}{2(2\pi)^3 z_h }$ is the identified-hadron phase space, and $X'$ denotes the final state excluding the identified hadron $h$. 
	
	We apply a transverse boost to transform the matrix element to the hadron frame~\cite{Collins:2011zzd}, defined by the conditions
	\bea 
	v_{pf}^+ = v_{hf}^+ \,, \qquad 
	\vec v_{T,pf} = \vec v_{T,hf} +   v_{hf}^+ \vec \beta_T \,, \qquad
	v_{pf}^-  
	= v_{hf}^- + 2 \vec \beta_T \cdot \vec v_{T,hf} +   {\vec \beta_T}^2 v_{hf}^+   \,, 
	\eea
	where $hf$ stands for the hadron frame in which the hadron transverse momentum $p_{h,T,hf}=0$, and $pf$ stands for the parton frame. 
	Here we have defined $ 
	\vec \beta_T \equiv  \vec p_{h,T,pf}/p_h^+ 
	$,  
	and we use the convention $ 
	v^\pm = v^0 \pm v^3  
	$. 
	This gives 
	\bea\label{eq:Mqt} 
	{\cal M}^{\rho\alpha}
	= &\; \sum_h Q_h \, \int d\Phi_h  \int  \d \xi \d^2{{x_T}} \,
	e^{i{\omega}\xi} 
	e^{ i \frac{\vec p_{hT}}{z_h} \cdot \vec x_T  } \, 
	\frac{p_{hT}^\alpha}{z_h} 
	\, \nonumber \\
	& \times
	{\rm Tr}\left[ \frac{i\sigma^{\rho+}\gamma_5}{2}
	\langle 0|  \chi_n(\xi, {x_T} )  
	\sumint_{X'} \, 
	|h(z_h,\vec 0), X',\text{out}\rangle \langle h(z_h,\vec 0) ,X',\text{out}|    
	{\bar \chi}_n(0) |0\rangle 
	\right]
	\nn \\ 
	= &\; \sum_h Q_h \, \int \frac{dz_h d^2 \vec k_T}{2 (2\pi)^3}\, (-k_T^\alpha) \,  z_h    \int  \d \xi \d^2{\vec x_T} \,
	e^{i{\omega}\xi} 
	e^{ - i \vec k_T \cdot \vec x_T  } \, 
	\, \nn \\ 
	&  \times     {\rm Tr}\left[ \frac{i\sigma^{\rho+}\gamma_5}{2}
	\langle 0|  \chi_n(\xi, {\vec x_T} )  
	\sumint_{X'} \, 
	|h(z_h,\vec 0), X',\text{out}\rangle \langle h(z_h,\vec 0) ,X',\text{out}|    
	{\bar \chi}_n(0) |0\rangle 
	\right] \,, 
	\eea 
	where $\vec k_T = - \vec p_{hT}/z_h$.
	Applying $- k_T^\alpha e^{ - i \vec k_T \cdot \vec x_T  } = -i{\partial}^\alpha_{T}    e^{ - i \vec k_T \cdot \vec x_T  }$ and integrating by parts, we obtain
	\bea  \label{eq:Mrhoalpha}   
	{\cal M}^{\rho\alpha}
	= \sum_h \int dz_h \, Q_h \,  z_h   \int  \frac{\d \xi }{4\pi} \,
	e^{i{\omega}\xi} 
	{\rm Tr}\left[ \frac{i\sigma^{\rho+}\gamma_5}{2}
	\langle 0| i{\partial}^\alpha_{T} \chi_n(\xi, \vec 0 )  
	\sumint_{X'} \, 
	|h(z_h,\vec 0), X',\text{out}\rangle \langle h(z_h,\vec 0) ,X',\text{out}|    
	{\bar \chi}_n(0) |0\rangle 
	\right]  \,,
	\eea 
	where we note that the sum over $h$, $z_h$, and $X'$ in
	this expression is precisely the spectral decomposition of the
	charge-flow operator $\widehat{\cal Q}(\hat n_0) \sim \sum_h \int dz_h Q_h z_h { \sum }_{X'} |h(z_h,\vec 0) X' \rangle \langle h(z_h,\vec 0) X' | $ along the $z$ direction, where $\hat n_0=(0,0,1)$. 
	Eq.~\eqref{eq:Mrhoalpha} eventually leads to the OPE of ${\cal M}^{\rho\alpha}$ 
	\bea 
	{\cal M}^{\rho\alpha}
	= \sum_h \int dz_h Q_h   H_h^{\rho\alpha}(z_h)  \,, 
	\eea 
	where 
	\bea \label{eq:Hope}
	H_h^{\rho\alpha}(z_h) = 
	z_h  \int  \frac{\d \xi }{4\pi} \,
	e^{i{\omega}\xi} \,
	{\rm Tr}\Bigg[& \frac{i\sigma^{\rho+}\gamma_5}{2}
	\langle 0| 
	\left( 
	iD^\alpha_T + g_s \int_\xi^\infty d\zeta F^{\alpha+}(\zeta)
	\right) 
	\chi_n(\xi, \vec 0 )  \notag \\
	&\times 
	\sumint_{X'} \, 
	|h(z_h,\vec 0), X',\text{out}\rangle \langle h(z_h,\vec 0) ,X',\text{out}|    
	{\bar \chi}_n(0) |0\rangle 
	\Bigg].
	\eea 
	Here we suppress the gauge links for simplicity and use the identity
	\bea 
	i\partial_T^\alpha \chi_n(\xi,\vec 0)
	=\left[ {\cal L}(\infty, \xi)iD_T^\alpha 
	+ g_s \int_\xi^\infty d\zeta {\cal L}(\infty,\zeta) F^{\alpha +}(\zeta) {\cal L}(\zeta,\xi)
	\right] \psi_n(\xi,\vec 0) \,, 
	\eea 
	for the gauge-invariant collinear field $\chi_n = {\cal L}(\infty,\xi)\psi_n(\xi,\vec 0) = {\cal P} \exp[-ig_s \int_\xi^\infty d\zeta A^+(\zeta) ] \psi_n(\xi,\vec 0)$. 
	
	Eq.~\eqref{eq:Hope} is the Collins first moment~\cite{Kang:2015msa} in the Trento convention, such that  $-1/2\, \epsilon_{T,\,\rho\alpha} \,H_h^{\rho \alpha} = M_h H_{1h/q}^{\perp(1)}$.
	The connection can also be derived directly from Eq.~\eqref{eq:Mqt}.  
	This relation identifies the coefficient of the linear term in $b_T$ as a charge-weighted sum of Collins moments. 
	The charge-flow OPE therefore manifests the Collins effect as a transverse charge dipole generated by the correlation between the transverse spin of the fragmenting quark and the transverse momenta of the charged hadrons. 
	
	Finally, one could perform the light-ray OPE by matching the light-ray operator \({\cal Q}\) onto the DGLAP operator and its transversity descendant in the hadron frame. We leave this extension to future work.

	\subsection*{SIDIS Phenomenology in COMPASS and EIC}
	We study transverse charge distributions in semi-inclusive deep-inelastic scattering (SIDIS) in the Breit frame, where \(\theta\) is the polar angle between the charge flow and the virtual photon. 
	The azimuthal angles \(\phi_s\) and \(\phi_n\) of the proton transverse spin and charge flow, respectively, are defined with respect to the reaction plane spanned by the incoming and outgoing leptons.
	
	\begin{figure}[!htb]
		\centering
		\includegraphics[width=0.99\linewidth]{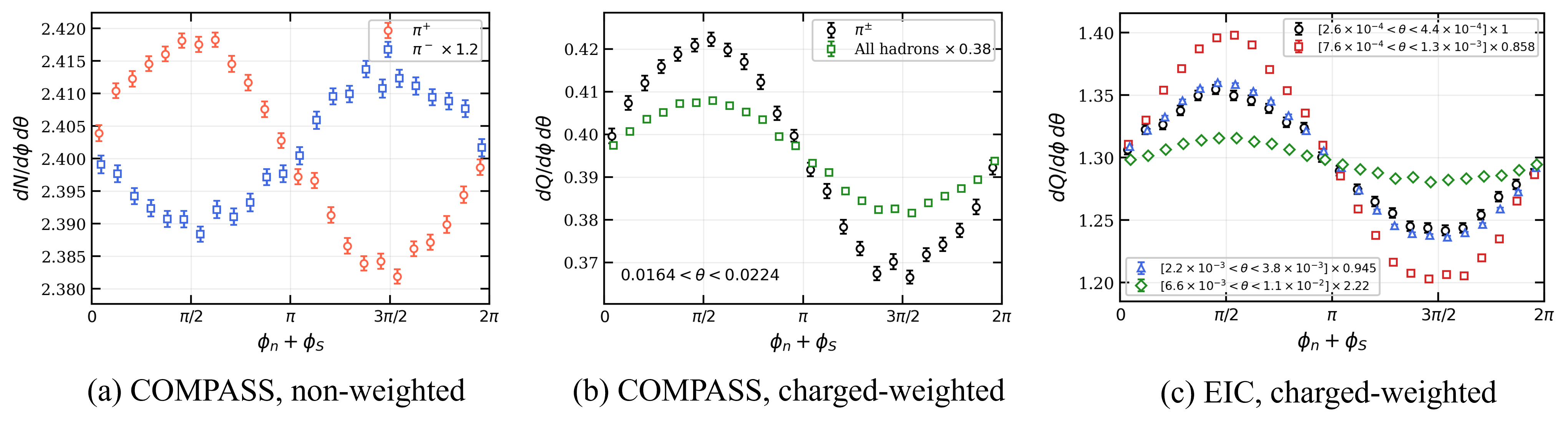}
		\caption{Azimuthal distributions of charged pions and all charged hadrons obtained from PYTHIA and StringSpinner simulations for COMPASS and EIC kinematics in the SIDIS Breit frame. Error bars represent statistical uncertainties.
			(a): unweighted \(\pi^+\) and \(\pi^-\) distributions in COMPASS kinematics, where the \(\pi^-\) distribution has been rescaled by a factor \(1.2\). 
			(b): charge-weighted \(\pi^\pm\) distribution and charge-weighted all-hadron distribution in COMPASS kinematics, with the latter rescaled by a factor \(0.38\).
			The displayed distributions correspond to the angular bin \(0.0164<\theta<0.0224\).
			(c): charge-weighted \(\pi^\pm\) distribution in EIC kinematics. The distributions in different $\theta$ bins have been rescaled to the same average value for better presentation.}
		\label{fig:SIDIS}
	\end{figure}
	
	We generate $N = 2.5 \times 10^{9}$ SIDIS events using PYTHIA~8.3~\cite{Bierlich:2022pfr} supplemented by StringSpinner~\cite{Kerbizi:2021pzn, Kerbizi:2022hhs, Kerbizi:2023cde, Kerbizi:2026iws}, which has been tuned to the COMPASS setup and data~\cite{COMPASS:2012ozz, COMPASS:2014ysd}.
	We apply the standard COMPASS kinematic cuts $Q^2>1~{\rm GeV}^2$, $W>5~{\rm GeV}$, $0.1<y<0.9$, and evaluate the azimuthal distribution of the charged pion multiplicities within the polar angle interval $0.0164 < \theta < 0.0224$, as shown in \cref{fig:SIDIS} (a) and (b). 
	Note that we have rescaled certain distributions for clarity, as indicated in the figure legends. 
	Both the unweighted pion distributions and the charge-weighted distributions exhibit a clear \(\sin(\phi_n+\phi_s)\) modulation. 
	The azimuthal asymmetry of the charge-weighted $\pi^\pm$ distribution is about an order of magnitude larger than those of the individual $\pi^+$ and $\pi^-$ distributions.
	We note that the all-hadron charge distribution in \cref{fig:SIDIS} (b) should be interpreted with caution, because the current version of StringSpinner includes only  pseudoscalar and vector mesons.
	Importantly, in the simulation, we observe that the charge-weighted \(\pi^\pm\) distribution does not suffer a loss of statistical precision relative to the separate \(\pi^+\) and \(\pi^-\) cases.
	
	In \cref{fig:SIDIS} (c), we repeat the simulation for EIC kinematics with \(E_e=10~{\rm GeV}\) and \(E_p=130~{\rm GeV}\) using $N = 2.5\times10^{11}$ events. 
	We use the same cuts as in the COMPASS study, except that we raise the virtuality cut to $Q^2>5~{\rm GeV}^2$ to suppress very small \(x\) events.
	Across the displayed \(\theta\) bins, the simulated charge distributions exhibit sizable \(\sin(\phi_n+\phi_s)\) modulations. Upon dividing out the SIDIS depolarization factor \(D_{NN}(y)=2(1-y)/[1+(1-y)^2]\)~\cite{Kerbizi:2023cde}, whose effective value over \(0.1<y<0.9\) is \(D_{NN}^{\rm eff}\simeq0.767\), the resulting asymmetries reach \(\sim10\%\).

\end{document}